\documentstyle[pra,aps]{revtex}
\input amssym.def
\input amssym.tex

\begin{document}

\draft
\title{The Exact Ground State of the Frenkel-Kontorova Model with Repeated 
Parabolic Potential\\
       II. Numerical Treatment}

\author{T. Scheidsteger}
\address{Hochschulrechenzentrum\\
Technische Hochschule Darmstadt\\
Petersenstr.30\\
D-64287 Darmstadt, Germany}
\author{H. Urbschat}  
\address{Fachbereich Physik and ICBM \\
         Carl von Ossietzky Universit\"at Oldenburg\\
         Postbox 2503 \\
         D-26111 Oldenburg,Germany}
\author{R. B. Griffiths}  
\address{Department of Physics \\
         Carnegie-Mellon University\\
         Pittsburgh, PA 15213, USA}
\author{H.J.Schellnhuber}
\address{Potsdam Institute for Climate Impact Research (PIK)\\
        P.O. Box 60 12 03 \\
	D-14412 Potsdam, Germany}
\date{Version of June 24, 1997}
\maketitle
\begin{abstract} 
	A procedure is described for efficiently finding the ground state
energy and configuration for a Frenkel-Kontorova model in a periodic potential,
consisting of $N$ parabolic segments of identical curvature in each period,
through a numerical solution of the convex minimization problem described in
the preceding paper.  The key elements are the use of subdifferentials to
describe the structure of the minimization problem; an intuitive picture of how
to solve it, based on motion of quasiparticles; and a fast linear
optimization method with a reduced memory requirement.  The procedure has been
tested for $N$ up to 200.
\end{abstract}  


\section{Introduction}
\label{s1}

	In the preceding paper \cite{GSU96}, the problem of finding ground
state energies and configurations for a Frenkel-Kontorova model in a periodic
potential formed by parabolic segments of identical (positive)
curvature, was reduced to that of minimizing a certain convex function over a
finite simplex.  While various aspects of the corresponding phase diagram in
the $N=2$ case can be worked out in a relatively straightforward manner
\cite{gri 90a,dl95,klt95}, minimizing the convex function for larger values of
$N$ represents a non-trivial problem in numerical analysis.  The basic reason
is that the convex function of interest does not have continuous derivatives,
and in the case of an irrational winding number, it possesses a dense set of
singularities.  Hence standard gradient methods run into difficulties.

	The approach we employ is based upon the concepts of subdifferential
and subgradient from the theory of convex functions \cite{rocka}, as explained
in Sec.~\ref{s2}.  The algorithm itself, described in detail in Sec.~\ref{s3},
is motivated by a physical model involving quasiparticles, Sec.~\ref{s3a}. 
An essential part of the procedure is a standard linear programming procedure
which we have simplified and adapted to the problem at hand, App.~A, so as to
speed it up substantially.  As a result, ground states for $N$ on the order of
100 are readily calculated, and larger values of $N$ are accessible with, of
course, a longer running time; see Sec.~\ref{s3d}.

\section{Minimization Using Subdifferentials}
\label{s2}

	As in part I, with some minor changes in notation, we assume the energy
per particle can be written in the form
\begin{equation}
\epsilon = \epsilon_0 + \epsilon_1,
\label{e2.1}
\end{equation}
where 
\begin{equation}
\epsilon_0=\sum_{j<k} \Delta t_j \Delta t_k {\cal G}(\zeta_{kj}),
\label{e2.2}
\end{equation}
\begin{equation}
\epsilon_1=\bar{\bf h}\cdot{\mbox{\boldmath $\psi$}}=-\bar{\mbox{\boldmath
$\eta$}}\cdot{\mbox{\boldmath $\zeta$}}= 
-\sum_{i=0}^{N-1}\bar\eta_i\zeta_i,
\label{e2.3}
\end{equation}
with $\bar h_0$ equal to zero,
\begin{equation}
\zeta_{kj}=-\zeta_{jk}=\zeta_k - \zeta_j = \sum_{i=j+1}^k \psi_i,
\label{e2.4}
\end{equation}
and
\begin{equation}
{\cal G}(\psi) = {\cal G}(-\psi) = {\cal G}(1-\psi) = 
\sum_{\nu=-\infty}^\infty w g({\psi + \nu \over w}).
\label{e2.5}
\end{equation}
The boldface letters denote $N$-component vectors, for example,
\begin{equation}
{\mbox{\boldmath $\zeta$}}= (\zeta_0, \zeta_1, \ldots \zeta_{N-1}).
\label{e2.6}
\end{equation}
A constant term independent of {\mbox{\boldmath $\zeta$}}\ has been
omitted from
(\ref{e2.2}), and
bars have been added to {\bf h}\ and {\mbox{\boldmath $\eta$}}\ in (\ref{e2.3})
to distinguish them
from quantities which we shall define later.  Note that
\begin{equation}
\sum_{i=0}^{N-1} \bar\eta_i = 0
\label{e2.7}
\end{equation}
because $\bar\eta_i=\bar h_{i+1}-\bar h_i$, and $\bar h_j$ is periodic in $j$, 
with period $N$.

	Our task is to find the {\mbox{\boldmath $\zeta$}}, or equivalently
{\mbox{\boldmath $\psi$}}, which minimize $\epsilon$ for a given
$\bar{\mbox{\boldmath $\eta$}}$, or equivalently $\bar{\bf h}$. The function
${\cal G}(\psi)$ is convex on the interval $0\leq\psi\leq 1$.  If its
derivative
\begin{equation}
{\cal B}(\psi)=d{\cal G}/d\psi=-{\cal B}(-\psi)={\cal B}(1+\psi)
\label{e2.8}
\end{equation}
were a continuous function, the minimum would satisfy the equation:
\begin{equation}
\bar{\mbox{\boldmath $\eta$}}={\mbox{\boldmath $\eta$}}
\label{e2.9}
\end{equation}
obtained by differentiating (\ref{e2.1}) ,
where the components of {\mbox{\boldmath $\eta$}}\ are given by
\begin{equation}
\eta_k=\sum_{j(\neq k)} \eta_{kj},
\label{e2.10}
\end{equation}
\begin{equation}
\eta_{kj}=\Delta t_k \Delta t_j \, \beta_{kj} = - \eta_{jk},
\label{e2.11}
\end{equation}
\begin{equation}
\beta_{kj} = {\cal B}(\zeta_{kj}) = - \beta_{jk}.
\label{e2.12}
\end{equation}

	But in fact ${\cal B}(\psi)$ has lots of discontinuities, see
Fig.~\ref{fg1},
and therefore we need some way to interpret the (formal) solution (\ref{e2.9})
in this case.  For this purpose it is convenient to employ the concepts of
subgradient and subdifferential as defined by Rockafellar \cite{rocka}.
Suppose that $F({\mbox{\boldmath $\zeta$}})$ is a real-valued function (which
need not be convex)
defined on some domain in ${\Bbb R}^N$.  We shall say that {\mbox{\boldmath
$\eta$}}\ is a {\it
subgradient} of $F$ at the point ${\mbox{\boldmath
$\zeta$}}=\bar{\mbox{\boldmath $\zeta$}}$ provided
\begin{equation}
F({\mbox{\boldmath $\zeta$}}) \geq F(\bar{\mbox{\boldmath
$\zeta$}})+{\mbox{\boldmath $\eta$}}\cdot ({\mbox{\boldmath
$\zeta$}}-\bar{\mbox{\boldmath $\zeta$}})
\label{e2.13}
\end{equation}
for all {\mbox{\boldmath $\zeta$}}\ where $F$ is defined.  The
collection of all
{\mbox{\boldmath $\eta$}}\ values for
which this inequality is satisfied for a given $\bar{\mbox{\boldmath $\zeta$}}$
is easily shown to
be a convex subset of ${\Bbb R}^N$, and is called the {\it subdifferential} of
$F$
at $\bar{\mbox{\boldmath $\zeta$}}$, denoted by $\partial F({\mbox{\boldmath
$\zeta$}})$.

	Given this definition, it is easy to show that $\epsilon$ in
(\ref{e2.1}) has a minimum at ${\mbox{\boldmath $\zeta$}}=\bar{\mbox{\boldmath
$\zeta$}}$ if and only if $\bar{\mbox{\boldmath $\eta$}}$ is
an element of $\partial\epsilon_0(\bar{\mbox{\boldmath $\zeta$}})$, the
subdifferential of
$\epsilon_0$ at $\bar{\mbox{\boldmath $\zeta$}}$.  In addition, the
subdifferential of a sum
($F_1+F_2+\cdots$) of convex functions is the sum of the subdifferentials of
the individual functions \cite{rockab}, understood as sums of sets of ${\Bbb
R}^N$; thus:
\begin{equation}
A + B = \{{\mbox{\boldmath $\zeta$}} + {\mbox{\boldmath $\zeta$}}' :
{\mbox{\boldmath $\zeta$}}\in A, {\mbox{\boldmath $\zeta$}}'\in B\},
\label{e2.14}
\end{equation}
together with its obvious generalization to the sum of three or more sets.

	These observations provide the key for interpreting equations
(\ref{e2.10}) to (\ref{e2.12}).  At some point $\psi'$ where ${\cal B}(\psi)$
is discontinuous, the subdifferential of ${\cal G}(\psi)$, for $\psi$ in the
range $0\leq\psi\leq 1$, consists of all points $\beta$ lying in the interval
\begin{equation}
{\cal B}_-(\psi')\leq \beta \leq {\cal B}_+(\psi')
\label{e2.15}
\end{equation}
where ${\cal B}_-(\psi')$ and ${\cal B}_+(\psi')$ are the left and right
derivatives of
${\cal G}(\psi)$ at $\psi'$, that is, the bottom and top of the
discontinuity in
the
graph of ${\cal B}$.  More generally, for $k > j$ we interpret $\beta_{kj}$ in
(\ref{e2.12}) as any point in the interval
\begin{equation}
{\cal B}_-(\zeta_{kj})\leq \beta_{kj} \leq {\cal B}_+(\zeta_{kj}),
\label{e2.16}
\end{equation}
where the lower limit is set equal to $-\infty$ if $\zeta_{kj}=0$, and the
upper limit is $+\infty$ if $\zeta_{kj}=1$, as a consequence of the constraints
\begin{equation}
\zeta_0 \leq \zeta_1 \leq \zeta_2 \leq \ldots \leq \zeta_{N-1} \leq \zeta_0+1.
\label{e2.17}
\end{equation}
For $j < k$, we define $\beta_{jk}=-\beta_{kj}$.  Of course, if ${\cal
B}(\psi)$
is
continuous at $\psi=\zeta_{kj}$, then $\beta_{kj}$ is the single point
${\cal B}(\zeta_{kj})$.

Consequently, $\partial\epsilon_0({\mbox{\boldmath $\zeta$}})$ is simply the
collection of all
${\mbox{\boldmath $\eta$}}$ obtained using (\ref{e2.10}), where for each $k >
j$, $\beta_{kj}$ in
(\ref{e2.12}) is allowed to vary over the interval (\ref{e2.16}), and for $j <
k$, $\beta_{jk}=-\beta_{kj}$.  Notice that this means that
$\eta_{jk}+\eta_{kj}$ is zero, and therefore
\begin{equation}
\sum_{i=0}^{N-1} \eta_i = 0
\label{e2.18}
\end{equation}
for any ${\mbox{\boldmath $\eta$}}$ in the subdifferential
$\partial\epsilon_0({\mbox{\boldmath $\zeta$}})$, which
corresponds to (\ref{e2.7}).  Note that the $\beta_{kj}$ are allowed to vary
{\it independently}, aside from the restriction $\beta_{jk}=-\beta_{kj}$.  As
there are thus $N(N-1)/2$ independent variables, some of which may be constant
because they do not correspond to discontinuities of ${\cal B}$,
$\partial\epsilon_0({\mbox{\boldmath $\zeta$}})$ is, in general, a fairly
complicated polyhedron,
of dimension less than or equal to $N-1$, the dimension of the space in ${\Bbb
R}^N$ satisfying the constraint (\ref{e2.18}).

	In the case $N=3$, the subdifferentials
$\partial\epsilon_0({\mbox{\boldmath $\zeta$}})$
are closed sets, either hexagons, lines, or points, depending upon the value of
${\mbox{\boldmath $\zeta$}}$.  Some of the lines and hexagons extend to
infinity.  A hexagon occurs
provided
\begin{equation}
\zeta_{10} = \mu_1 w - \nu_1,\ \ \zeta_{20} = \mu_2 w - \nu_2,
\label{e2.19}
\end{equation}  
where $\mu_1$, $\mu_2$, $\nu_1$, and $\nu_2$ are integers, in which case
$\zeta_{10}$ and $\zeta_{20}$, as well as $\zeta_{21}=\zeta_{20}-\zeta_{10}$,
are at discontinuities of ${\cal B}$.  A line occurs when ${\cal B}$ is
discontinuous at
one of the three values $\zeta_{10}$, $\zeta_{20}$, or $\zeta_{21}$, but not at
the other two, and $\partial\epsilon_0({\mbox{\boldmath $\zeta$}})$ is a point
if ${\cal B}$ is
continuous at all three values.  If $w$ is irrational, the discontinuities of
${\cal B}(\psi)$ are a dense set in $\psi$, and consequently the
subdifferentials of
$\epsilon_0$ for different ${\mbox{\boldmath $\zeta$}}$ have no points in
common.  If $w$ is
rational, adjacent hexagons overlap at their common edges and vertices, and
each edge and each vertex is itself a subdifferential of $\epsilon_0$ for a
range of ${\mbox{\boldmath $\zeta$}}$ values.  For both rational and irrational
$w$, the hexagons
cover the entire plane satisfying the constraint (\ref{e2.18}), with the
exception, when $w$ is irrational, of a set of zero measure.  A similar comment
applies to larger values of $N$, and therefore in numerical studies it suffices
to consider the $N-1$ dimensional polyhedra obtained when every $\zeta_{kj}$
falls on a discontinuity of ${\cal B}$.

	It is sometimes helpful to think of the collection of subdifferentials
$\partial\epsilon_0({\mbox{\boldmath $\zeta$}})$ as ${\mbox{\boldmath
$\zeta$}}$
varies as generated by placing a set
of $N(N-1)/2$ points on the graph of ${\cal B}(\psi)$ at positions
$(\zeta_{kj},\beta_{kj})$.  Note that the $\zeta_{kj}$ cannot be varied
independently, as they are determined by a set of $N-1$ parameters, see
(\ref{e2.4}).  However, each of the $N(N-1)/2$ points on the graph can be moved
independently in the vertical direction, as long as it is on a discontinuity of
${\cal B}$, to form the collection of $\beta_{kj}$ values which generate the
subdifferential for a fixed ${\mbox{\boldmath $\zeta$}}$.
 
\section{Numerical Procedure}
\label{s3}

\subsection{Introduction}
\label{s3a}
	The problem of finding the ${\mbox{\boldmath $\zeta$}}$ which minimizes
$\epsilon$,
(\ref{e2.1}) , for a given ${\mbox{\boldmath $\eta$}}$ is equivalent, as noted
in Sec.~\ref{s2}
above, to finding the ${\mbox{\boldmath $\zeta$}}$ such that
$\bar{\mbox{\boldmath $\eta$}}$ falls in the
subdifferential $\partial\epsilon_0({\mbox{\boldmath $\zeta$}})$.  Furthermore,
the $N-1$
dimensional polyhedra which arise when all the $\zeta_{kj}$ fall at the
discontinuities of ${\cal B}(\psi)$ fill up the relevant $N-1$ dimensional
hyperplane
(\ref{e2.18}) \ except for a set of measure zero, and as we assume that
${\mbox{\boldmath $\eta$}}$
is only specified with some limited numerical precision, we can in practice
limit ourselves to a consideration of such polyhedra.

	The general idea of the algorithm is as follows.  Starting from some
${\mbox{\boldmath $\zeta$}}$ with all $\zeta_{kj}$ at discontinuities of ${\cal
B}$, test whether the
target $\bar{\mbox{\boldmath $\eta$}}$ lies inside
$\partial\epsilon_0({\mbox{\boldmath $\zeta$}})$. If it does, the
problem has been solved.  If it does not, use the information obtained from the
test in order to choose a new {\mbox{\boldmath $\zeta$}}\ closer to the desired
value, and repeat
the test.  The test itself, steps 3 and 5 in the algorithm as summarized below,
involves a linear optimization procedure with an execution time which
(typically) varies as $N^3$, which is relatively expensive when $N$ is large.
Consequently, the test is preceded in our algorithm by various steps whose aim
is to provide, with a relatively small number of operations, a value of
${\mbox{\boldmath $\zeta$}}$ close to the final solution.

	To begin with, we replace the actual ${\cal B}(\psi)$ with an
approximate,
piecewise constant function ${\cal B}^*(\psi)$ which has discontinuities at the
points
\begin{equation}
\psi=\mu w - \nu,
\label{e3.1}
\end{equation}  
where $\mu$ and $\nu$ are integers, and 
\begin{equation}
|\mu| \leq M
\label{e3.2}
\end{equation} 
for some finite bound $M$, which can be increased later if necessary.  This is
a sensible procedure, because the size of the discontinuities decreases
exponentially with $|\mu|$.  Between two successive discontinuities $\psi'$ and
$\psi''$, the function ${\cal B}^*$ is defined to be a constant lying halfway
between
${\cal B}_+(\psi')$ and ${\cal B}_-(\psi'')$, see Fig.~\ref{fg1}. 
Consequently,
the
discontinuities of ${\cal B}^*$ are somewhat larger than those of the exact
${\cal B}$, and
(\ref{e2.16}) is replaced by:
\begin{equation}
{\cal B}^*_-(\zeta_{kj})\leq \beta_{kj} \leq {\cal B}^*_+(\zeta_{kj}).
\label{e3.3}
\end{equation} 

	Note that if 
\begin{equation}
w=p/q
\label{e3.4}
\end{equation}
is a rational number, with $p$ and $q$ relatively prime positive integers, the 
discontinuities of ${\cal B}$, (\ref{e3.1}), are the points
\begin{equation}
\psi = s/q,
\label{e3.5}
\end{equation}
where $s$ is any integer (and may have factors in common with $q$).  The
definition of ${\cal B}^*$ is the same as before; though it should be
noted that
the
discontinuity interval of ${\cal B}$ at a point (\ref{e3.5}) is made up of
contributions from an infinite number of discontinuities from derivatives of
terms on the right side of (\ref{e2.5}).  Of course, if $M$ in
(\ref{e3.2})
is equal to $q-1$ (or larger), ${\cal B}^*$ and ${\cal B}$ are identical, and
step 5 can be 
eliminated
from the algorithm described below.

	In order to motivate the initial steps in the algorithm, it is helpful
to think of $\zeta_0, \zeta_1,\ldots$ as representing the positions of a set of
$N$ quasiparticles located on a circle of unit circumference, Fig.~\ref{fg2},
and subjected to two kinds of forces, corresponding to $\epsilon_0$ and
$\epsilon_1$, thought of as potential energies.  In this picture, $\bar\eta_k$
represents an external force exerted on particle $k$, and $\eta_{kj}$,
(\ref{e2.11}), the force which particle $k$ exerts on particle $j$.  The
minimization condition (\ref{e2.9}) can then be interpreted as stating that,
for every $k$, the external force exerted on particle $k$ is equal to the sum
$\eta_k$ of the forces which it exerts on the other particles, which is the
same as saying that the net force on particle $k$ is zero.  The pair force
$\eta_{kj}$ is constant as long as $\zeta_{kj}$ is not at a discontinuity of
${\cal B}^*$, while if it is at such a discontinuity, it can take any
value, see
(\ref{e2.11}), corresponding to $\beta_{kj}$ in the range (\ref{e3.3}).  In
addition, there is a hard core interaction which prevents two quasiparticles
from passing through each other, and ensures that the inequalities
(\ref{e2.17}) are satisfied.  Note that there can very well be solutions to the
minimization problem in which some of these inequalities are equalities.  If,
for example, $\zeta_2=\zeta_3$, then $\beta_{32}$ can be very large and
negative, see the remarks following (\ref{e2.16}), corresponding to the fact
that the hard core allows particle 3 to exert a very large (negative) force on
particle 2.

	The initial steps of the algorithm consist of a number of horizontal
and vertical shifts; the terminology comes from the picture of points on the
graph of ${\cal B}$, Sec.~\ref{s2}.  A {\it horizontal shift} is a
change in the
positions of the quasiparticles, and thus the $\zeta_{kj}$, with the
$\beta_{kj}$ (and thus the $\eta_{kj}$) held fixed, while a {\it vertical
shift} is a change in the set of $\beta_{kj}$ values with the quasiparticle
positions, and thus the $\zeta_{kj}$, held fixed.  In addition, we shall make
use of the concept of a {\it cluster}, which means a collection of
quasiparticles, with their labels belonging to an index set $J$ containing
$|J|$ members, with the property that the collection is connected by a set of
``pair bonds'' $(kj)$, $k$ and $j$ members of $J$, with $\zeta_{kj}$ at one of
the discontinuities of ${\cal B}^*$.  For example, if $\zeta_{21}$ and
$\zeta_{42}$
fall on discontinuities of ${\cal B}^*$, then it is possible, but not
necessary,
to
define a cluster $J=\{1,2,4\}$ of $|J|=3$ quasiparticles, as in Fig.~\ref{fg2}.
We shall always think of the entire collection of quasiparticles as divided up
among a set of mutually disjoint clusters, where a quasiparticle which does not
belong to a larger cluster constitutes its own cluster containing only one
element.  If all the quasiparticles belong to a single cluster of $|J|=N$
elements, we shall call this a {\it complete cluster}. 	In any horizontal
shift, the clusters are moved rigidly, in the sense that $\zeta_{jk}$ does not
change if $j$ and $k$ belong to the same cluster.  (This must obviously be the
case whenever the cluster is linked together by bonds for which the
$\beta_{kj}$ fall in the interiors of the corresponding discontinuity intervals
(\ref{e3.3}).)  Conversely, a vertical shift is always applied to a single
cluster.

\subsection{Summary of the Algorithm}
\label{s3b}

	The algorithm for finding a minimum consists of the following steps,
details of which are given below in Sec.~\ref{s3c}.

	0. Initialization:
	
	Choose an initial approximate ${\cal B}^*$ by, for example, setting
$M=4$ in
(\ref{e3.2}), and some initial values for the $\zeta_i$ satisfying
(\ref{e2.17}),
with a set of clusters specified (e.g., each quasiparticle might belong to its
own cluster).

	1. Horizontal shift I:
	
	Calculate a ``velocity'' for each cluster, and use this to carry out a
horizontal shift until for the first time some $\zeta_{kj}$ for $j$ and $k$
belonging to different clusters reaches a discontinuity of ${\cal B}^*$, in
which
case we shall say that these two clusters have ``collided'' to form a temporary
combined cluster.  Go to step 2.

	2. Vertical shift:
	
	Carry out a vertical shift on the temporary combined cluster following
the prescription given in (\ref{e3.8}) below and in the remarks which follow,
and apply the test which is described there.  If the result of the test is
negative, the combined cluster is rejected, the collection of clusters is
defined to be the same as it was before the collision, and the algorithm
returns to step 1.  If the result of the test is positive, the temporary
combined cluster becomes permanent, and is considered part of the collection of
clusters for the next step in the algorithm.  If this cluster is complete, go
to step 3; if not, return to step 1.

	3. Linear optimization I:
	
	With all the quasiparticles in a single cluster, apply linear
optimization, as discussed in Sec.~\ref{s3c} below, to produce a vertical shift
which maximizes a non-negative parameter $\lambda$.  If $\lambda\geq 1$, then
the target $\bar{\mbox{\boldmath $\eta$}}$ is inside the polyhedron
$\partial\epsilon_0({\mbox{\boldmath $\zeta$}})$
associated with the current ${\mbox{\boldmath $\zeta$}}$ in the
approximation in
which ${\cal B}$ has
been replaced by ${\cal B}^*$; go to step 5.  If $\lambda < 1$, then
$\bar{\mbox{\boldmath $\eta$}}$ is
not inside the polyhedron; go to step 4.

	4. Horizontal shift II:
	
	Use the $\beta_{kj}$ resulting from the linear optimization in 3 to
divide the collection of quasiparticles into two clusters, which undergo a
horizontal shift relative to each other until they collide (as in step 1) to
form a new, combined cluster which is a complete cluster.  Return to step 3.

	5. Linear optimization II:
	
	Repeat the linear optimization step 4, but with each $\beta_{kj}$ now
restricted to the corresponding {\it exact} interval (\ref{e2.16}).  If,
however,
$\zeta_{kj}$ falls at a point where ${\cal B}^*$, unlike ${\cal B}$, has no
discontinuity,
then the corresponding $\beta_{kj}$ is placed at the center of the
corresponding discontinuity of ${\cal B}$, and is treated as a constant, not a
variable, during the linear optimization.  If the optimization yields
$\lambda\geq 1$, the current ${\mbox{\boldmath $\zeta$}}$ is the desired
solution to the
minimization problem, and the algorithm stops.  If $\lambda$ is less than 1,
${\cal B}^*$ is replaced by another approximation to ${\cal B}$ constructed in
the same
way, but using a larger value of $M$ in (\ref{e3.2}).  The current
${\mbox{\boldmath $\zeta$}}$ values
are changed by very small amounts so that none of the $\zeta_{kj}$ fall at
discontinuities of the new ${\cal B}^*$, and the algorithm returns to step 1.

\subsection{Details of the Algorithm}
\label{s3c}

	The explanations given below are numbered in the same way as the steps
in the preceding summary.

	1.  Given a set of $\eta_{kj}$ values, the net force on the $k$'th
quasiparticle is
\begin{equation}
\bar\eta_k - \eta_k = \bar\eta_k - \sum_{j(\neq k)} \eta_{kj}.
\label{e3.6}
\end{equation} 
Were the force given by a continuous function, it would be possible to find the
energy minimum by assigning to each quasiparticle a velocity proportional to
the force acting on it, and then solving the resulting dynamics.  What is
actually done in the algorithm is to assign to each cluster a velocity given by
the total force acting on all the quasiparticles in the cluster divided by the
number of particles in the cluster, which is the average force per particle:
\begin{equation}
v_J= (1/|J|) \sum_{k\in J} (\bar\eta_k - \eta_k).
\label{e3.7}
\end{equation}
It is only the relative cluster velocities which are of interest in determining
the horizontal shift; adding the same constant to every $v_J$ will make no
difference, and one can arrange (for example) that the cluster containing
$\zeta_0$ remains fixed.  The clusters are then shifted by amounts proportional
to their respective velocities until the first ``collision'' occurs, in the
sense that $\zeta_{kj}$ for $k$ in one cluster and $j$ in another reaches a
discontinuity of ${\cal B}^*$.

	2.  Once the temporary, combined cluster $J_c$ has been formed, a
vertical shift is applied to the $\eta_{kj}$ for $k$ and $j$ in $J_c$.  This is
done by first calculating a preliminary value for the shift of the $\eta_{kj}$
or $\beta_{kj}$ values from the formula
\begin{equation}
\delta\eta_{kj} = \Delta t_k \Delta t_j \, \delta\beta_{kj} = 
	[(\bar\eta_k - \eta_k) - (\bar\eta_j - \eta_j)] / |J_c|.
\label{e3.8}
\end{equation}
The motivation for this choice is the following.  If all the quasiparticles
were in a single cluster, $|J_c|=N$, the change (\ref{e3.8}) would result in a
new set of pair forces
\begin{equation}
\eta'_{kj} = \Delta t_k \Delta t_j \, \beta'_{kj} = \eta_{kj} +\delta\eta_{kj}
\label{e3.9}
\end{equation}
with the property that 
\begin{equation}
\bar\eta_k=\eta'_k = \sum_{j(\neq k)}\eta'_{kj},
\label{e3.10}
\end{equation}
that is, one would have solved the minimization problem.  With $|J_c| < N$, the
result would, instead, be to make the difference $\bar\eta_k-\eta'_k$, the sum
of the forces acting on quasiparticle k, independent of $k$ for all $k$ in
$J_c$, and to minimize
\begin{equation}
\sum_{k\in J_c} (\bar\eta_k-\eta'_k)^2
\label{e3.11}
\end{equation} 
as much as is possible by changing only the  pair interactions $\eta_{kj}$ 
inside the cluster $J_c$. 

	However, formula (\ref{e3.8}) does not take account of the possibility
that the $\beta'_{kj}$ in (\ref{e3.9}) might lie outside the interval
(\ref{e3.3}) determined by ${\cal B}^*$.  When this is the case, the new
$\beta'_{kj}$ is placed at whichever end of the discontinuity interval lies
closest to the value given by (\ref{e3.9}).  The test for rejecting or
retaining
the combined cluster $J_c$ is then the following.  If for each of the pairs $k$
and $j$ for which $k$ belongs to one of the clusters involved in the collision
and $j$ to the other, the new $\beta'_{kj}$ is at one of the ends of the
interval (\ref{e3.3}), the combined cluster $J_c$ is rejected, whereas if at
least one of these values falls in the interior of the corresponding interval,
$J_c$ is accepted.  Note that whether the cluster $J_c$ is accepted or
rejected, the new values of $\beta'_{kj}$, and thus the corresponding
$\eta'_{kj}$, produced in the vertical shift are retained when going on to the
next step of the algorithm, which is either step 1 or, in the case in which
$J_c$ is accepted and $|J_c|=N$, step 3.

	The algorithm would still function correctly if a combined cluster were
never rejected.  However, this would mean having to apply linear optimization,
step 3, more often, and would result in a slower computation.  The process of
allowing clusters to move relative to each other past discontinuities which
represent relatively small changes compared to the large forces representing a
situation ``far from equilibrium'' helps to achieve a better preliminary value
of ${\mbox{\boldmath $\zeta$}}$ before going on to step 3.

	3, 4.  The linear optimization step is basically a test to see whether
the target $\bar{\mbox{\boldmath $\eta$}}$ lies inside the polyhedron
representing
$\partial\epsilon_0({\mbox{\boldmath $\zeta$}})$ in the approximation in which
${\cal B}$ is replaced by
${\cal B}^*$.  The idea is to begin at a particular vertex ${\mbox{\boldmath
$\eta$}}^c$ of the
polyhedron, Fig.~\ref{fg3}, and draw a straight line from ${\mbox{\boldmath
$\eta$}}^c$ to the
target $\bar{\mbox{\boldmath $\eta$}}$.  Points along this line are of the form
\begin{equation}
{\mbox{\boldmath $\eta$}} ={\mbox{\boldmath $\eta$}}^c + \lambda
(\bar{\mbox{\boldmath $\eta$}} - {\mbox{\boldmath $\eta$}}^c),
\label{e3.12}
\end{equation}
where $\lambda$ is a number between 0 and 1.  The linear optimization
procedure, the details of which are given in the appendix, determines the
largest value of $\lambda$ for which a point of the form (\ref{e3.12}) lies
inside or on the boundary of the polyhedron.  If this value is less than 1, as
in Fig.~\ref{fg3}, the target lies outside the polyhedron, and the point
(\ref{e3.12})  determined by the maximum value of $\lambda$ specifies a facet
of the polyhedron lying in the direction of the target, as viewed from the
starting vertex ${\mbox{\boldmath $\eta$}}^c$.  This facet is generated by
letting ($N-2$) of the
$\beta_{kj}$ vary over their entire discontinuity intervals, while the
remaining $\beta_{kj}$ are fixed either at the top or at the bottom of their
discontinuity intervals. The $\zeta_{kj}$ corresponding to the former are
``rigid'' in the sense that they cannot be altered by a horizontal shift
(which, by definition, must leave the $\beta_{kj}$ unchanged), and one can
identify two clusters of quasiparticles, each one connected by such rigid
bonds.

	Once these two clusters have been identified, they can be shifted
relative to each other, in a direction which is obvious, until they collide at
the first discontinuity of ${\cal B}^*$.  This collision results in a new,
complete
cluster, and the corresponding $\partial\epsilon_0({\mbox{\boldmath $\zeta$}})$
is a polyhedron
adjacent to the one considered earlier, and shares with it the facet which was
identified in the previous linear optimization step.

	Note that this new cluster is accepted without carrying out the test
used in step 2 of the algorithm.  Also, in the unlikely event that the maximum
$\lambda$ corresponds to the intersection of two or more facets of the
polyhedron, the actual optimization algorithm described in the appendix will,
in effect, ``choose'' one of these facets, and thus the polyhedron adjacent to
it in the direction of the target $\bar{\mbox{\boldmath $\eta$}}$.

	5. If the optimization carried out in step 3 yields $\lambda \geq 1$,
the target $\bar{\mbox{\boldmath $\eta$}}$ lies inside the polyhedron generated
by the
discontinuities of ${\cal B}^*$, but it may or may not lie inside the
corresponding
polyhedron generated by restricting the $\beta_{kj}$ to lie in the exact
interval (\ref{e2.16}) \ for the corresponding discontinuity.  To test whether
this
is the case, one repeats the linear optimization technique of step 3, but now
starting with each $\beta_{kj}$ at its {\it exact} maximum possible value, and
constrained to be greater than or equal to its {\it exact} minimum possible
value, with the exception of those $\beta_{kj}$ for which $\zeta_{kj}$ does not
fall at a discontinuity of ${\cal B}^*$, which are assigned fixed values at the
center of the appropriate (exact) discontinuity intervals of ${\cal B}$.  One
could,
of course, make {\it all} of the $\beta_{kj}$ variable during the linear
optimization process. However, as the discontinuity intervals in ${\cal B}$
which are
not in ${\cal B}^*$ are, by construction, relatively small, the main effect of
using
a larger set of variables would be to slow down the linear optimization without
much hope of actually finding a solution with $\lambda \geq 1$ when the
restricted search yields one with $\lambda < 1$.  Note that if such a
restriction does result in overlooking a $\lambda \geq 1$ solution, this
solution, or one equivalent to it, will nevertheless be found later when the
number of discontinuities of ${\cal B}^*$ is increased.

  If the maximum value of $\lambda$ obtained by using linear optimization with
the exact discontinuity intervals of ${\cal B}$ is still greater than or equal
to 1,
then the current ${\mbox{\boldmath $\zeta$}}$, and thus the current set of
$\zeta_{kj}$, represents
an actual solution to the minimization problem, and this does not depend upon
the approximations used in constructing ${\cal B}^*$, because the resulting
$\beta_{kj}$ values all fall within the range where ${\cal B}^*$ is
identical to
${\cal B}$.

	If, on the other hand, one finds that $\lambda$ is less than 1, this
means that the current ${\mbox{\boldmath $\zeta$}}$ is not a solution to the
true minimization
problem; instead, it is as good as one can do using the approximate ${\cal
B}^*$.  To
do better, it is necessary to increase the number of discontinuities.  Our
procedure at this point is to throw away all the information associated with
the $\beta_{kj}$ values obtained in the immediately preceding step of linear
optimization, and simply start over again at step 1 using the current
${\mbox{\boldmath $\zeta$}}$.
One might be able to improve the algorithm in this respect, but since the
initial steps of the algorithm are relatively fast, it does not seem likely
that one would obtain a significant increase in speed.

\subsection{Implementation and Performance}
 \label{s3d}

	The program we constructed to implement the algorithm described above
was tested in the following way.  We chose a winding number equal to the
inverse golden mean (0.618\dots) and a value of $\kappa$, see \cite{GSU96}, of
0.6, resulting in a set of 60 pairs ($\psi$ and $1-\psi$) of discontinuities of
${\cal B}$ in the interval $0\leq\psi\leq 1$ with a magnitude greater than a
resolution of $10^{-20}$.  (Some tests used alternative values for $\kappa$,
0.1, 0.5, and 1.0, for which there are 146, 66, and 47 pairs of
discontinuities, respectively, exceeding this resolution.)  The parabolas were
assumed to be equally spaced, with $\Delta t_l$ independent of $l$. Then we
employed the following ``inverse strategy''.  With $N$ fixed, random values of
$\zeta_{j}$, lying on the full set of discontinuities of ${\cal B}$ were
chosen, subject to the constraints (\ref{e2.17}), and values $\eta_{kj}$ inside
the discontinuity intervals were also chosen randomly, thus defining---see
(\ref{e2.10}), (\ref{e2.9}), and (\ref{e2.3})---$\bar{\mbox{\boldmath $\eta$}}$
for a model of $N$
parabolas with the solution to its energy minimization problem already known.
The algorithm was then applied to this model starting at the initialization
step 0, with a (different) random collection of $\zeta_{i}$, and a choice of
$M$ (the number of pairs of discontinuities in the approximate ${\cal B}^*$) to
search for the correct solution.

	Note that the running time increases linearly with $M$.  But since as
$M$ increases, the size of the discontinuities in ${\cal B}^*$ is decreasing,
the probability that a random choice of $\bar{\mbox{\boldmath $\eta$}}$ will
actually require a
larger value of $M$ goes to zero exponentially with increasing $M$.  We found
that using an initial value of $M=2$ when $N$ is small saves a lot of running
time, and in almost all cases $M=10$ was sufficient to find a ground state with
$N$ up to 100.

	To determine the $N$ dependence of the running time, we generated and
timed 10 distinct potentials for each $N$ in an increasing series up to
$N=115$.  Using an HP 9000 model 735 workstation with 64 MB of RAM and a CPU
with 20 MFLOPs, the average time in seconds required to find a solution was
approximately $10^{-5}N^4$, or a quarter of an hour for $N=100$.  The time
required for linear optimization varies as $N^3$, but as $N$ becomes larger,
``quasiparticle dynamics'' takes up a larger fraction of the time: 60-70\% for
$N=100$.  The algorithm found the correct ground state in particular cases
for $N$ as large as 200.

\section*{ Acknowledgments }
We would like to thank S. Aubry, P. Delaly, A. Hamm, R. S. MacKay and P. Rujan
for useful discussions.  This work was supported by the Deutsche
Forschungsgemeinschaft (Bonn, Germany).

\appendix
\section{Linear Optimization}

	The linear optimization procedure used in steps 3 and 5 of the
algorithm functions in the following way. Written out in terms of components,
Eq.(\ref{e3.12}) has the form:
\begin{equation}
\eta_k=\eta_k^c+\lambda(\bar\eta_k-\eta_k^c)=
	\sum_{j\neq k} \Delta t_k \Delta t_j  \beta_{kj}
\label{eA.1}
\end{equation}
for $k=1,2,\ldots,N-1$.
The equation for $k=0$ can be omitted because of the constraints (\ref{e2.7})
and
(\ref{e2.18}). In turn, the $\beta_{kj}$ for $k>j$ can be written as
\begin{equation}
\beta_{kj}={\cal B}_+^*(\zeta_{kj}) + x_{kj}[{\cal B}_-^*(\zeta_{kj})-{\cal
B}_+^*(\zeta_{kj})],
\label{eA.2}
\end{equation}
where the $N(N-1)/2$ variables $x_{kj}$ take values on the closed interval
\begin{equation}
0\leq x_{kj}\leq 1.
\label{eA.3}
\end{equation}
In step 5 of the algorithm, ${\cal B}^*$ in (\ref{eA.2}) should be
replaced with
the
exact function ${\cal B}$.  The vertex ${\mbox{\boldmath $\eta$}}^c$ of the
polyhedron is defined as the
value of ${\mbox{\boldmath $\eta$}}$ when $\lambda=0$, and is obtained by
setting all the $x_{kj}$
in (\ref{eA.2}) equal to zero, and inserting the resulting $\beta_{kj}$ in the
right side of (\ref{eA.1}). It is, thus, a known quantity, as is also the
target $\bar{\mbox{\boldmath $\eta$}}$.  By contrast, the $x_{kj}$ and
$\lambda$
are considered as
variables subject to the $N-1$ equalities obtained by inserting (\ref{eA.2}) in
(\ref{eA.1}), and the inequalities (\ref{eA.3}), along with $\lambda \geq 0$.
If we define the objective function to be equal to $\lambda$, maximizing
$\lambda$ subject to these constraints becomes a standard problem in the theory
of linear optimization. For a compact but clear description, see \cite{num
rec}.  The usual approach is to replace the second set of inequalities in
(\ref{eA.3}) by the requirement that the ``slack'' variables
\begin{equation}
y_{kj}=1-x_{kj}
\label{eA.4}
\end{equation}
be non-negative. However, in order to minimize the number of variables, we have
modified the formulation in \cite{num rec}, as discussed below, so that at any
point in the calculation, either $x_{kj}$ or $y_{kj}$, but not both, appears as
a variable in the tableau.

	The first task is to construct a tableau in restricted normal form, in
which the $N-1$ equalities (\ref{eA.1}), expressed in terms of the $x_{kj}$
using (\ref{eA.2}), are transformed into expressions in which $N-1$ of the
$x_{kj}$ variables are expressed as linear functions of the remaining $x$'s and
$\lambda$.  To see how this is done, note that since the $k=0$ equation is
missing from the set (\ref{eA.1}), $x_{k0}$ occurs only in the $k$'th equation,
and in none of the others. Therefore, if, for each $k>0$, the coefficient of
$x_{k0}$ in (\ref{eA.2}) is non-zero, we can rewrite the $N-1$ equalities in
the form
\begin{equation}
x_{k0}=\ldots
\label{eA.5}
\end{equation}
for $k=1,2,\ldots,N-1$, with no $x_{j0}$ appearing anywhere on the right
side of
any of these equations.

	It may, however, happen that the coefficient of some $x_{k0}$ in
(\ref{eA.2}) vanishes, because ${\cal B}^*(\psi)$ does not have a discontinuity
at $\psi=\zeta_{k0}$.  In this case the following strategy can be employed.
Construct a graph in which the $N$ quasiparticles are vertices, and they are
connected together in one cluster using $N-1$ edges $(k,j)$ chosen so that for
each edge the coefficient of $x_{kj}$ in (\ref{eA.2}) does not vanish. Such a
graph is a tree with no closed loops, because a connected graph of $N$ vertices
which includes a loop will have at least $N$ edges.  One then chooses some
vertex $m$ which has only one edge connecting it to another vertex $n$, and
solves the equation (\ref{eA.1}) with $k=m$ for $x_{mn}$ (or $x_{nm}$) in terms
of the other $x$'s.  After this, eliminate vertex $m$ and the edge $(mn)$ from
the graph, and repeat the process. At each stage one finds (it is useful to
work out an example) that the only $x$'s appearing on the right side of the
equation either do not correspond to edges of the original graph, or else to
edges which have already been eliminated. The latter, however, can be expressed
(recursively) as functions of the $x$'s corresponding to edges which were not
present in the original graph, thus leading to a set of $N-1$ equalities of a
form similar to (\ref{eA.5}). Note that the $x_{kj}$ whose coefficients are
zero in (\ref{eA.2}) can be completely eliminated in constructing the linear
optimization tableau, as they play no role.

	Once the tableau has been constructed for the restricted normal form,
the optimization is carried out as indicated in \cite{num rec}, with the
following difference.  When considering increasing the variable associated with
a particular column, it is necessary to take account of both positive and
negative entries in the column in order to test whether one of the left
$x_{kj}$ will reach either the upper or lower limit in (\ref{eA.3}). If the
choice is to set a left variable $x_{kj}$ equal to 1, it is replaced as a right
variable by the corresponding $y_{kj}$, which is 0, and the tableau is
appropriately transformed.

\begin{figure} 
\caption{
	The exact ${\cal B}(\psi)$ (solid lines) for a
winding number equal to the inverse golden mean (.618\dots) and  $\kappa =
0.1$ is shown by solid lines, and the  approximation ${\cal B}^*(\psi)$, which
retains only the largest discontinuities (see text) by dashed lines.
}
\label{fg1}
\end{figure}

\begin{figure} 
\caption{
An example of $N=6$ quasiparticles lying on a circle of unit
circumference and partially coupled to a cluster with index set $J= \{1,2,4\}$.
}
\label{fg2}
\end{figure}

\begin{figure} 
\caption{
Schematic representation of the linear optimization procedure, see
(\ref{e3.12}).  Points ${\mbox{\boldmath $\eta$}}$ with $\lambda$ less than 0.6
lie inside the
polyhedron (in this case a polygon), so that the target $\bar{\mbox{\boldmath
$\eta$}}$ lies outside
it, and the open circle corresponding to the maximum value of $\lambda$
indicates the starting point for the next step of the algorithm.
  }
\label{fg3}
\end{figure}

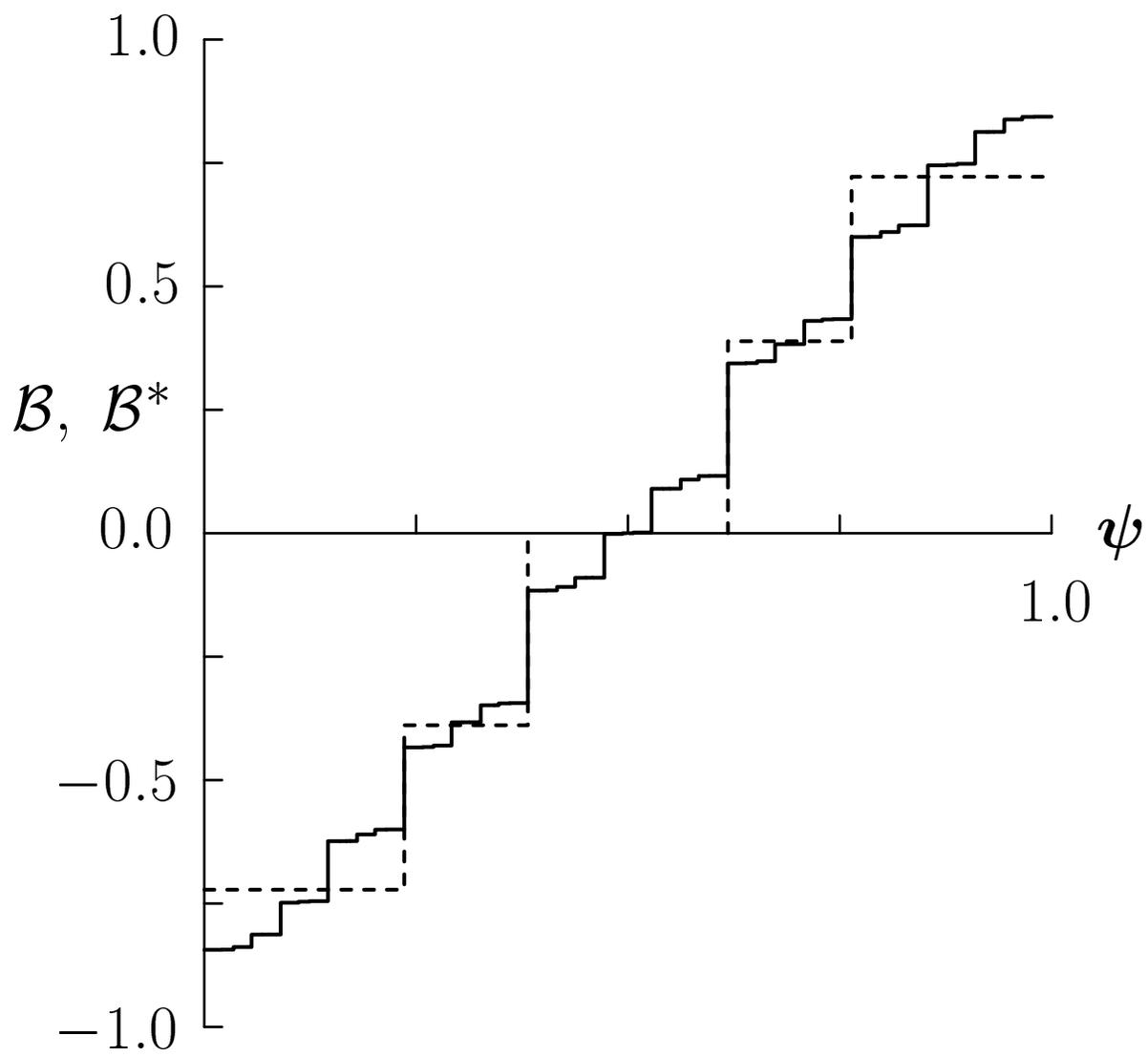
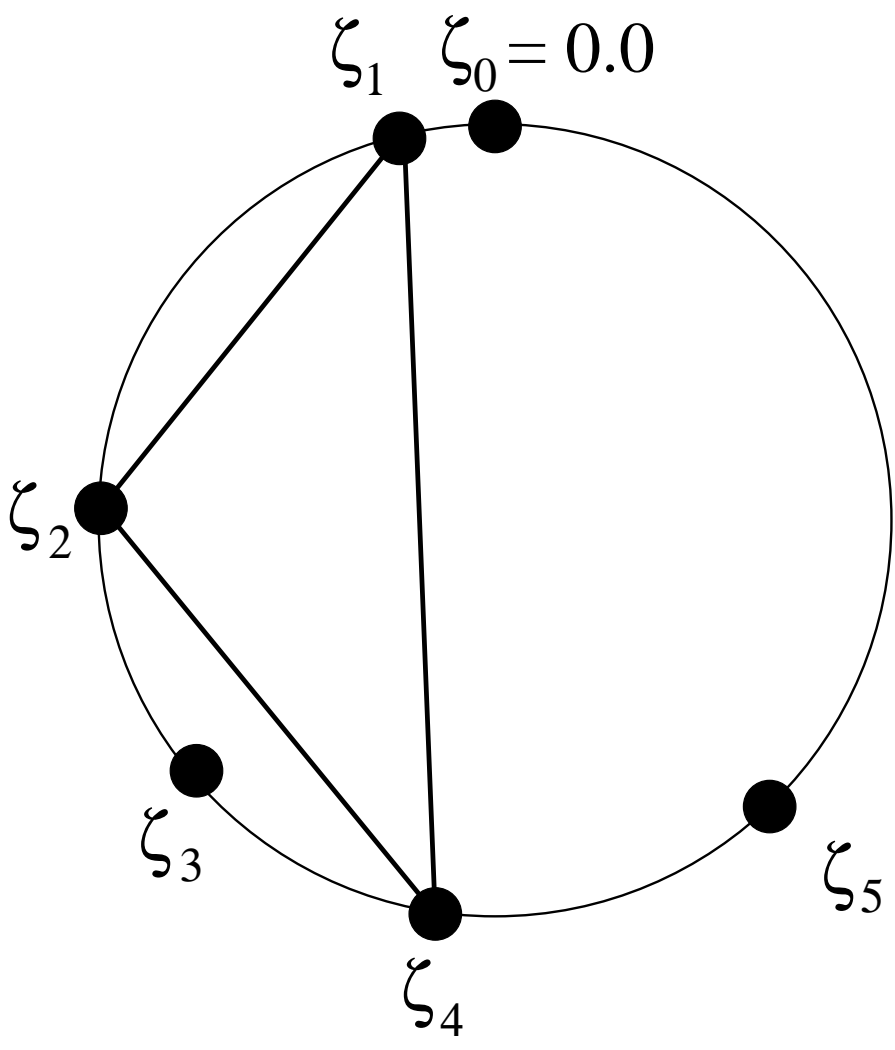
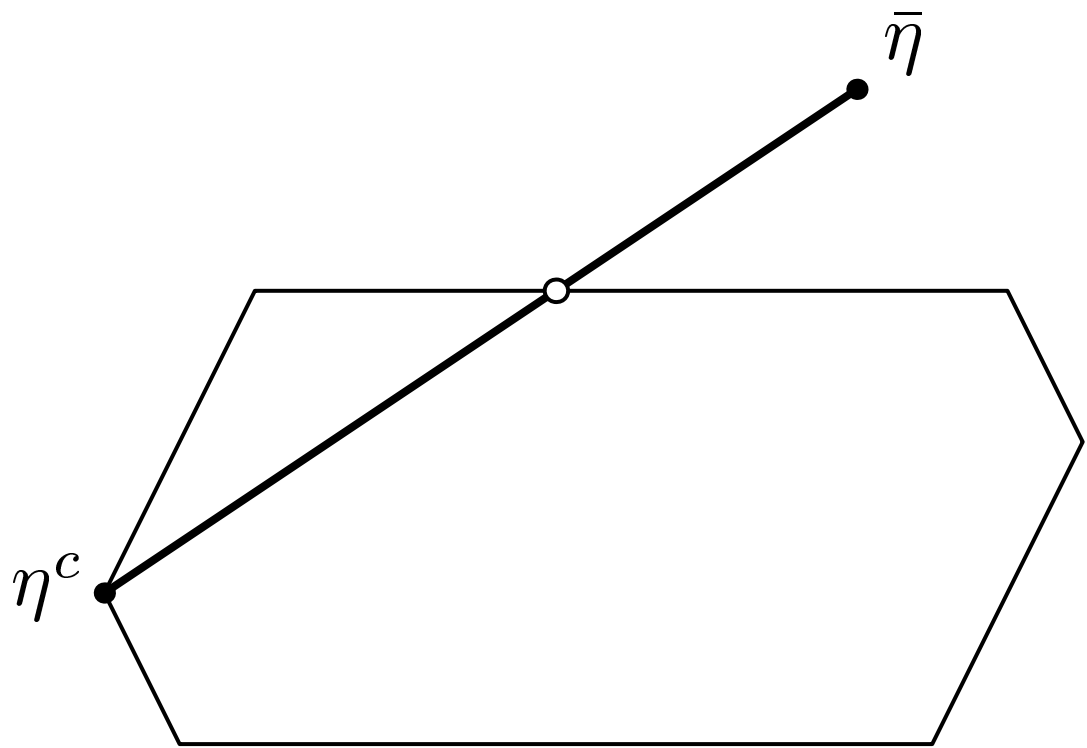
\end{document}